\documentclass[a4paper]{article}

\usepackage{INTERSPEECH2022}

\usepackage{multirow}
\usepackage{CJKutf8}
\usepackage{url}
\usepackage{hyperref}

\title{
End-to-end Speech-to-Punctuated-Text Recognition
}
\name{Jumon Nozaki$^1$, Tatsuya Kawahara$^1$, Kenkichi Ishizuka$^2$, Taiichi Hashimoto$^2$}
\address{
  $^1$ Graduate School of Informatics, Kyoto University, Japan\\
  $^2$RevComm, Inc., Japan
}
\email{\{nozaki, kawahara\}@sap.ist.i.kyoto-u.ac.jp\\
\{ishizuka, taiichi.hashimoto\}@revcomm.co.jp}

\begin{document}

\maketitle

\begin{abstract}
Conventional automatic speech recognition systems do not produce punctuation marks which are important for the readability of the speech recognition results.
They are also needed for subsequent natural language processing tasks such as machine translation.
There have been a lot of works on punctuation prediction models that insert punctuation marks into speech recognition results as post-processing.
However, these studies do not utilize acoustic information for punctuation prediction and are directly affected by speech recognition errors.
In this study, we propose an end-to-end model that takes speech as input and outputs punctuated texts.
This model is expected to predict punctuation robustly against speech recognition errors while using acoustic information.
We also propose to incorporate an auxiliary loss to train the model using the output of the intermediate layer and unpunctuated texts.
Through experiments, we compare the performance of the proposed model to that of a cascaded system.
The proposed model achieves higher punctuation prediction accuracy than the cascaded system without sacrificing the speech recognition error rate.
It is also demonstrated that the multi-task learning using the intermediate output against the unpunctuated text is effective.
Moreover, the proposed model has only about 1/7th of the parameters compared to the cascaded system.
\end{abstract}

\section{Introduction}
Automatic speech recognition (ASR) systems have made significant progress with advances in deep neural networks (DNNs)~\cite{bahdanau2016end,karita2019comparative,baevski2020wav2vec}.
ASR systems are widely used in many applications, such as conversational robots and speech captioning systems.
However, conventional ASR systems do not generate punctuation marks, which affect the readability of the transcripts.
Also, unpunctuated texts are not suitable for subsequent applications of natural language processing.
For example, some previous work has shown that unpunctuated texts degrade the performance of machine translation and named entity recognition~\cite{cho2012segmentation, peitz2011modeling, nguyen2020improving}.

To address this problem, research on punctuation prediction models has been conducted.
The punctuation prediction model is used to insert punctuation marks into speech recognition results as post-processing.
Before DNNs became widely used, n-gram language models~\cite{gravano2009restoring}, support vector machines~\cite{akita2006sentence}, conditional random fields~\cite{lu2010better, akita2011automatic} were mainly used to predict punctuation.
After that, using DNN models such as LSTM~\cite{hochreiter1997long} and Transformer~\cite{vaswani2017attention} were investigated to output punctuated sentences from unpunctuated word sequences as input~\cite{tilk2015lstm, yi2019self}.
More recently, there have been a lot of works on using pre-trained models based on the Transformer architecture such as BERT~\cite{devlin2019bert} for punctuation prediction~\cite{makhija2019transfer, yi2020adversarial}.
Thanks to the powerful Transformer architecture pre-trained with a huge amount of text corpora, they achieved state-of-the-art performance on the IWSLT2011 dataset~\cite{che2016punctuation}, a well-known benchmark for punctuation prediction.
Then, the focus of research has shifted towards using more advanced pre-trained models such as RoBERTa~\cite{liu2019roberta} and ELECTRA~\cite{clark2019electra} and more advanced training techniques to further push the performance of punctuation prediction~\cite{courtland2020efficient,chen21d_interspeech}.

These conventional studies implicitly assume a cascaded application of two separate models: an ASR model and a punctuation prediction model.
However, there are some disadvantages caused by the nature of the cascaded system.
First, these previous studies generally use only lexical features and not acoustic (prosodic) features although acoustic information such as pauses and pitches are considered to be important for punctuation prediction.
Ignoring acoustic information for punctuation prediction also makes the system vulnerable to speech recognition errors.
Second, recent studies use pre-trained models such as BERT, but they have a large number of parameters and are not suitable for on-device systems.
Because much attention has been given to running ASR systems on mobile devices such as smartphones and tablets~\cite{he2019streaming}, it is important to develop a fast and lightweight model that can run on limited computational resources.
Lastly, there is accuracy degradation due to segmentation errors.
When feeding an ASR output into the punctuation model based on a pre-trained model such as BERT, it needs to be tokenized according to the vocabulary of the pre-trained model, but tokenizing texts without punctuation and with ASR errors is difficult, thus causing some tokenization errors.
One might be tempted to use the same vocabulary for the ASR model and the pre-trained model to make tokenization unnecessary, but the vocabulary of the pre-trained model is not suitable for the vocabulary of ASR because it is case-sensitive and contains a lot of unpronounceable symbols such as parentheses.
This issue of the segmentation error is more serious for languages without explicit word boundaries (e.g., Japanese and Chinese).

In this study, we propose an end-to-end model for speech-to-punctuated-text recognition.
Specifically, we use the stacked Transformer encoder layers and train them with CTC loss~\cite{graves2006connectionist} using speech as input and punctuated texts as output.
This model predicts punctuation robustly against ASR errors and segmentation errors while using acoustic information.
We also propose a method to train the model using an auxiliary loss calculated from the output of the intermediate layer and unpunctuated texts, in addition to the original loss in the last layer.
The experiments are conducted on English and Japanese datasets to demonstrate the effectiveness of the proposed model compared to the cascaded system.

\section{Related Work}

\subsection{Punctuation prediction using acoustic features}
There are studies that use acoustic features to predict punctuation~\cite{christensen2001punctuation, sunkara2020multimodal}.
In these studies, it is assumed that there is a separately trained ASR model that outputs unpunctuated texts.
For each token in the ASR output, the corresponding acoustic features are obtained, which are used as input to train a punctuation prediction model.
While they still use separate models for ASR and punctuation prediction, we propose an end-to-end model from speech to punctuated text, which is optimized as a single model.

\subsection{Auxiliary loss in intermediate layers}
Some studies investigated using an auxiliary loss for the output of intermediate layers to train CTC-based ASR~\cite{fernandez2007sequence,sanabria2018hierarchical,lee2021intermediate}.
They proposed to use phones~\cite{fernandez2007sequence}, a small-sized vocabulary~\cite{sanabria2018hierarchical}, or even the same label as that for the last layer~\cite{lee2021intermediate} for the output of the intermediate layer.
Our method can be seen as an extension of these studies to speech-to-punctuated-text recognition, in that we use unpunctuated texts for the output of the middle layer and punctuated texts for the output of the last layer.

\subsection{End-to-end approach}
Some previous work proposed to train an end-to-end model that takes speech as input and outputs ``formatted'' texts.
Caseiro et al. \cite{Caseiro2020} proposed a system that takes speech as input and outputs a case-sensitive text directly in an end-to-end manner.
The most similar work to ours is that of Mimura et al.~\cite{mimura2021endtoend}.
They proposed an end-to-end model that takes speech as input and outputs a clean text with punctuation and without disfluency for the Japanese parliamentary meetings.
While they addressed inserting punctuation, removing fillers, and substituting colloquial expressions at the same time, we focus on punctuation because punctuation is ubiquitous in any style of speech and plays an essential role to improve the readability of text.
We also propose an auxiliary loss that is specific to punctuation.

\section{Speech-to-Punctuated-Text Recognition}
\subsection{Task Definition}
In this study, we define speech-to-punctuated-text recognition as the problem of outputting a token sequence containing punctuation marks $\mathbf{y}_{\text{pnct}}$ from an acoustic feature sequence $\mathbf{X}$ as input.
For punctuation marks in English, we consider a comma (,), a period (.), and a question mark (?).

\subsection{Baseline Model}
\begin{figure}[t]
\centering
\includegraphics[width=\columnwidth]{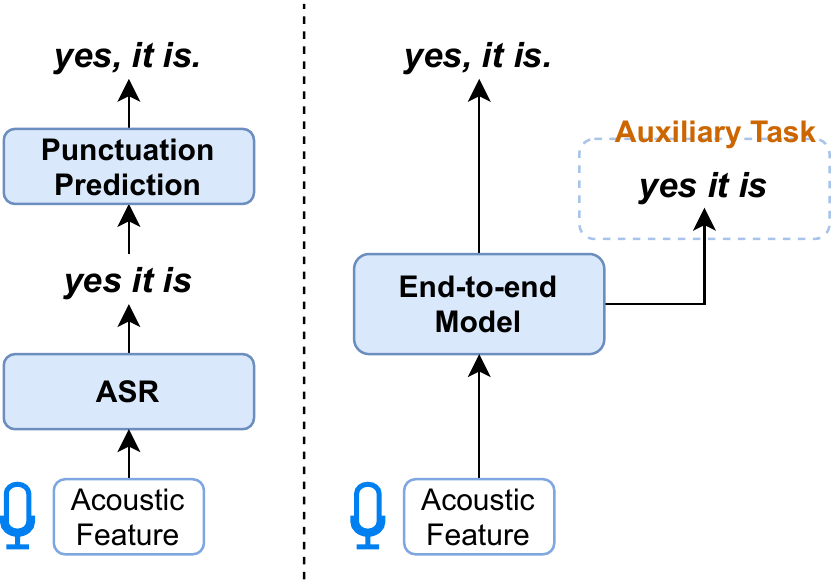}
\caption{Overview of the cascaded system (left) and the proposed end-to-end model (right) for speech-to-punctuated-text recognition.}
\label{fig:model}
\end{figure}

The mainstream method for speech-to-punctuated-text recognition is to train a speech recognition model and a punctuation prediction model separately, and then cascade them together for inference.
An overview of the cascaded system is shown on the left-hand side of Figure~\ref{fig:model}.
In this study, we use a Transformer-based ASR model and BERT-based punctuation prediction model as a baseline system.
Specifically, the ASR output is first tokenized according to the vocabulary of BERT.
Then, BERT is fine-tuned with a task to classify each token according to the type of punctuation marks inserted immediately after the token.
In the case of English data, each token is assigned to the following classes: ``O'', ``COMMA'', ``PERIOD'', and ``QUESTION'', where ``O'' indicates there is no punctuation after the token.
For example, suppose that the input sequence is ``yes it is'', the corresponding label sequence is ``COMMA, O, PERIOD'' since the properly punctuated version of the input text is ``yes, it is.''
Using the classification task into these punctuation classes, BERT is fine-tuned with the following cross-entropy loss.
\begin{equation}
    \mathcal{L}_{\mathrm{CE}}=-\sum_{k=1}^{K} t_{k} \log p_{k}
\end{equation}
where $K$ is the number of classes (in this case $K = 4$), $p_k$ is the predicted probability for label $k$, and $t_k$ is the target probability for label $k$.
We use the one-hot label for $t_k$: $t_k = 1$ if k is the corresponding label, else $t_k = 0$.

\section{Proposed Method}
In this study, we propose a model that directly outputs a token sequence containing punctuation marks $\mathbf{y}_{\text {pnct}}$ from an acoustic feature sequence $\mathbf{X}$ in an end-to-end manner.
An overview of the proposed model is shown on the right-hand side of Figure~\ref{fig:model}.

We use stacked Transformer Encoder layers as a model architecture and train it with a CTC loss function.
The CTC loss function calculates the sum of the probabilities of the alignments which can be reduced to the output label series $\mathbf{y}$, as represented by the following equation.
\begin{equation}
    P_{\mathrm{CTC}}(\mathbf{y} \mid \mathbf{X})= \sum_{\mathbf{a} \in \Gamma(\mathbf{y})} P(\mathbf{a} \mid \mathbf{X})
\end{equation}
Here, $\Gamma^{-1}(\boldsymbol{a})$ is a function that concatenates consecutive identical tokens and removes special blank tokens.
The alignment probability $P(\mathbf{a} \mid \mathbf{X})$ is formulated under the conditional independence assumption between tokens:
\begin{equation}
    P(\mathbf{a} \mid \mathbf{X})= \prod_{s} P(a_{s} \mid \mathbf{X})
\end{equation}
where $a_s$ represents the $s$-th symbol of $\mathbf{a}$, and $P(a_{s} \mid \mathbf{X})$ represents the probability of observing $a_{s}$ at time s.

The proposed model uses an $l$-layer Transformer encoder and is trained to minimize the following CTC loss function for the output of the final $l$-th layer $\mathbf{X}_{l}$.

\begin{align}
\label{lossctc}
\mathcal{L}_{\mathrm{CTC}} = -\log P_{\mathrm{CTC}}(\mathbf{y}_{\text {pnct}} \mid \mathbf{X}_{l})
\end{align}

Although Eq.~(\ref{lossctc}) is sufficient to train the model in an end-to-end manner, we also propose to use an auxiliary CTC loss calculated with the token sequence without punctuation $\mathbf{y }_{\text {unpnct}}$ and the output of the intermediate layer to more effectively train the model.
The loss for the output of the middle $\lfloor l / 2 \rfloor$-th layer $\mathbf{X}_{\lfloor l / 2 \rfloor}$ is calculated as follows.

\begin{align}
\label{lossinter}
\mathcal{L}_{\mathrm{inter}} = -\log P_{\mathrm{CTC}}(\mathbf{y}_{\text {unpnct}} \mid \mathbf{X}_{\lfloor l / 2 \rfloor})
\end{align}

The final loss function is formulated as a weighted linear sum of the Eq.~(\ref{lossctc}) and Eq.~(\ref{lossinter}).

\begin{align}
\label{losstotal}
\mathcal{L}_{\text {total}}=\lambda_{\text {CTC}} \mathcal{L}_{\text {CTC}} + \lambda_{\text {inter}} \mathcal{L}_{\text {inter}}
\end{align}
where $\lambda_{\text {CTC}}$ and $\lambda_{\text {inter}}$ are the coefficients of the respective loss terms.

During inference, we do not make the predictions in the middle layer, but only the predictions in the final layer, thus there is no overhead in inference time.

\section{Experimental Evaluations}
\subsection{Datasets}
We used two datasets of different languages: English and Japanese.

The MuST-C corpus~\cite{di2019must} was used as the English dataset.
Although MuST-C is a dataset mainly used for research on speech translation, in this study, we extracted the English speech and the English script with punctuation from the English-German speech translation data and used them as paired data.
Comma (,), period (.), and a question mark (?) are considered to be punctuation marks.
We used the ``tst-COMMON'' set as a test set.
As a preprocessing, the scripts were all converted to lowercase.
Note that in the MuST-C corpus, periods and question marks are mostly at the end of an utterance, so the prediction of these marks is relatively easy.

JCALL is an in-house Japanese dataset, which consists of audio recordings of conversations between a salesperson and a customer in inside sales and those between an operator and a customer in call centers.
We considered the Japanese comma (\begin{CJK}{UTF8}{ipxm}、\end{CJK}), the Japanese full-stop mark (\begin{CJK}{UTF8}{ipxm}。\end{CJK}), and a question mark (?) as punctuation marks.
Utterances were segmented using VAD, and punctuation marks are present in the middle of an utterance as well as at the end of an utterance.

For both datasets, we removed utterances longer than 30 seconds from the training set due to computational resource constraints.
We also removed some duplicate utterances from the training set so that the number of utterances with the same reference text was at most 300.
The statistics of the two datasets are shown in Table~\ref{tab:data}.

\subsection{Experimental Setup}
For the baseline cascaded system, we trained an ASR model and a punctuation prediction model separately.
The ASR model consisted of stacked Transformer encoder layers and was trained with the CTC loss.
During training, speech and unpunctuated texts were used as paired data.
The number of Transformer layers was 12, the dimension of the hidden layer was 256, and the number of heads was 4.
For input features, 80-dimensional log-mel spectrum features were used, and SpecAugment~\cite{park2019specaugment} was used for data augmentation.
For the vocabulary of ASR, we used 2000 tokens created by SentencePiece~\cite{kudo2018sentencepiece} for MuST-C, and 1,923 characters for JCALL.
For the punctuation prediction model, we used the pre-trained \texttt{base}-sized BERT model taken from Hugging Face’s \texttt{transformers} package~\cite{wolf-etal-2020-transformers}\footnote{Publicly available at \url{https://huggingface.co/}. We used
\href{https://huggingface.co/bert-base-uncased}{bert-base-uncased} for MuST-C and \href{https://huggingface.co/cl-tohoku/bert-base-japanese-whole-word-masking}{cl-tohoku/bert-base-japanese-whole-word-masking} for JCALL.
}.
We added a linear layer to the final layer of the BERT model and fine-tuned it with the punctuation classification task.
The transcripts of the training set of MuST-C and JCALL were used as the fine-tuning data of BERT.
The training was done for 10 epochs on a single GPU using Adam~\cite{kingma2015adam} as the optimizer with a learning rate of $10^{-5}$.
The batch size was set to 32.

The proposed end-to-end model was trained with the same architecture and the training method as the ASR model of the cascaded system, except that punctuated texts instead of unpunctuated texts were used as labels for training.

For the proposed model and the ASR model of the cascaded system, we conducted training with and without using the auxiliary loss represented by Eq.~(\ref{lossinter}).
When using the auxiliary loss, $\lambda_{\text {CTC}}$ and $\lambda_{\text {inter}}$ in Eq.~(\ref{losstotal}) were equally set to 0.5.

\begin{table}[t]
\centering
\caption{The number of utterances and punctuation marks for each split of MuST-C and JCALL datasets.
``\begin{CJK}{UTF8}{ipxm}、\end{CJK}'' and ``\begin{CJK}{UTF8}{ipxm}。\end{CJK}'' denote the Japanese comma and the Japanese period, respectively.
}
\begin{tabular}{llrrrr}
\hline
\multirow{2}{*}{Dataset} & \multirow{2}{*}{Split} & \multirow{2}{*}{\#Utterances} & \multicolumn{3}{c}{\#Punctuation} \\
& &  & , (\begin{CJK}{UTF8}{ipxm}、\end{CJK}) & . (\begin{CJK}{UTF8}{ipxm}。\end{CJK}) & ? \\
\hline
\multirow{3}{*}{MuST-C} & train & 225K & 287K & 246K & 21K \\
& dev & 1,423 & 2,117 & 1,599 & 158 \\
& test & 2,641 & 2,761 & 2,804 & 239 \\
\hline
\multirow{3}{*}{JCALL} & train & 150K & 211K & 150K & 25K \\
& dev & 4,000 & 4,247 & 3,353 & 473 \\
& test & 4,000 & 4,043 & 3,365 & 451 \\
\hline
\end{tabular}
\label{tab:data}
\end{table}

\begin{table*}[t]
\centering
\caption{Evaluation results on MuST-C and JCALL datasets.}
\begin{tabular}{l|cccccc|cccccc}
\hline
& & \multicolumn{4}{c}{MuST-C} & & & \multicolumn{4}{c}{JCALL} &  \\
\multirow{2}{*}{Model} & WER\footnotemark & \multicolumn{4}{c}{Punctuation F1-Score (\%) $\uparrow$} & \#Params & CER & \multicolumn{4}{c}{Punctuation F1-Score (\%) $\uparrow$} & \#Params \\
&  (\%) $\downarrow$ & , & . & ? & avg & ($M$) $\downarrow$ & (\%) $\downarrow$  & \begin{CJK}{UTF8}{ipxm}、\end{CJK} & \begin{CJK}{UTF8}{ipxm}。\end{CJK} & ? & avg & ($M$) $\downarrow$ \\ 
\hline
Cascaded System & 21.2 & 63.3 & 88.5 & 70.5 & 74.1 & 128 & 14.5 & 49.0 & 63.6 & 60.0 & 57.5 & 126 \\
\quad + intermediate loss & 20.0 & \bf{64.5} & 89.1 & 69.7 & 74.4 & 128 & \bf{14.3} & 49.2 & 63.2 & 59.6 & 57.3 & 126 \\
\hline
End-to-End (\bf{Proposed}) & 27.4 & 55.4 & 91.7 & 56.8 & 68.0 & \bf{18} & \bf{14.3} & \bf{59.2} & \bf{69.1} & \bf{74.2} & \bf{67.4} & \bf{18} \\
\quad + intermediate loss & \bf{19.8} & 61.1 & \bf{92.5} & \bf{74.1} & \bf{75.9} & \bf{18} & \bf{14.3} & 56.9 & 68.7 & 74.0 & 66.5 & \bf{18} \\
\hline
\end{tabular}
\label{tab:main}
\end{table*}

\subsection{Evaluation}
As a measure of the accuracy of ASR, we used Character Error Rate (CER) for JCALL and Word Error Rate (WER) for MuST-C.
All punctuation marks in the output texts were removed before the calculation.
As a measure of the accuracy of the punctuation prediction, the F1 score for each type of punctuation mark and its average were calculated.
As the output of the model contains ASR errors, it is not possible to simply calculate the F1 score by comparing it to the ground-truth.
Therefore, we first aligned the predicted text with the ground-truth text, then calculated the F1 score. 
In addition, we compared the total number of parameters in each model.

\subsection{Results}
The results of the experiments are shown in Table~\ref{tab:main}.

In the experiment on MuST-C, when we did not use the intermediate loss, the WER of the proposed model was largely worse than that of the cascaded system.
However, when the intermediate loss was incorporated, the WER was improved to a comparable level to the cascaded system, showing the effectiveness of using the intermediate loss.
We conjecture that simply training the mapping from speech to punctuated texts was difficult, but using the auxiliary loss worked as a good regularization and stabilized the training.
In terms of punctuation prediction accuracy, the proposed model was more accurate than the cascade system by an absolute 1.5\% average F1 score (75.9\% vs. 74.4\%).

In the experiments on JCALL, the proposed model obtained higher punctuation prediction accuracy (67.4\% vs. 57.5\%) while maintaining comparable CER (14.3\%) as the cascade system.
We suspect that the lower punctuation prediction accuracy of the cascade system is due to the segmentation errors.
Since the Japanese language does not use a space as word separators, the tokenization of unpunctuated texts with ASR errors is more likely to be erroneous compared to the English dataset.
On the other hand, the proposed method can predict punctuated texts in an end-to-end manner without the need for tokenization, which results in robust punctuation prediction.
When trained with the auxiliary intermediate loss, we did not see an improvement of CER and punctuation prediction accuracy in this case.  

For experiments on both datasets, the proposed model required only about 1/7th  of the parameters compared with the cascade system.

\footnotetext[2]{The WER of the baseline using the attention-based encoder-decoder model reported in the MuST-C paper~\cite{di2019must} was 27.0\%.}

\begin{table}[t]
\centering
\caption{Word error rates (WER) and averaged punctuation F1 scores when different labels are used for the last layer and the middle layer on MuST-C.
``-'' indicates that intermediate loss was not used.
The last row shows the result of multitask learning for the output of the last layer.
``E2E'' stands for end-to-end.}
\begin{tabular}{ccccc}
\hline
\multirow{2}{*}{Last layer} & \multirow{2}{*}{Middle layer} & \multirow{2}{*}{E2E?} & WER & F1 \\
 &  &  & (\%) $\downarrow$ & (\%) $\uparrow$ \\
\hline
$\mathbf{y}_{\text {pnct}}$ & - & \checkmark & 27.4 & 68.0 \\
$\mathbf{y}_{\text {unpnct}}$ & - &  & 21.2 & 74.1 \\
$\mathbf{y}_{\text {unpnct}}$ & $\mathbf{y}_{\text {unpnct}}$ &  & 20.0 & 74.4 \\
$\mathbf{y}_{\text {pnct}}$ & $\mathbf{y}_{\text {pnct}}$ & \checkmark & 25.2 & 72.4 \\
$\mathbf{y}_{\text {pnct}}$ & $\mathbf{y}_{\text {unpnct}}$ & \checkmark & \bf{19.8} & \bf{75.9} \\
$\mathbf{y}_{\text {pnct}}$ \& $\mathbf{y}_{\text {unpnct}}$ & - & \checkmark & 20.4 & 30.0 \\
\hline
\end{tabular}
\label{tab:layer-analysis}
\end{table}

\subsection{Analysis}
To analyze the effectiveness of using different labels for the last layer and the middle layer, we trained models using different settings on MuST-C.
We used either the punctuated text $\mathbf{y}_{\text {pnct}}$ or the unpunctuated text $\mathbf{y}_{\text {unpnct}}$ as the label for the last layer, and used either or neither of them as the label for the middle layer.
For a reference, we also trained a model using multitask learning of $\mathbf{y}_{\text {pnct}}$ and $\mathbf{y}_{\text {unpnct}}$ for the output of the last layer.
For the models trained to predict $\mathbf{y}_{\text {unpnct}}$ in the last layer, we used the BERT-based punctuation prediction model to insert punctuation and report the F1 score.
Table~\ref{tab:layer-analysis} shows the result of the experiments.
We can see that using punctuated texts for the last layer and unpunctuated texts for the middle layer achieved the best WER (19.8\%) and F1 punctuation score (75.9\%).
Training the model with punctuated texts for both of the last layer and the middle layer significantly degraded WER (25.2\%), which confirms using the unpunctuated texts for the middle layer was essential.
We conjecture that this was because gradually making the task difficult as going to the deeper layer made the training stable.
When we trained a model using $\mathbf{y}_{\text {pnct}}$ and $\mathbf{y}_{\text {unpnct}}$ in the last layer, the accuracy of punctuation prediction was drastically degraded because it was not effectively trained.

\section{Conclusions}
In this study, we proposed an end-to-end model that directly predicts a punctuated text using speech as input.
The proposed model can utilize acoustic information for punctuation prediction and can be robust against ASR errors and segmentation errors.
The evaluation experiments showed that the proposed model can achieve higher recognition accuracy with much fewer parameters than the conventional cascaded system.
In addition, we showed the effectiveness of using an auxiliary loss using unpunctuated texts for the output of the intermediate layer.
Our approach also has advantages in terms of inference speed and the simplicity of its architecture.
Our study sets out a future research direction of using an end-to-end model for speech-to-punctuated-text recognition.
In the future, we plan to study the improvement of the proposed model using different architectures and its application to real-time speech recognition.

\bibliographystyle{IEEEtran}

\bibliography{mybib}

\begin{thebibliography}{10}
\providecommand{\url}[1]{#1}
\csname url@samestyle\endcsname
\providecommand{\newblock}{\relax}
\providecommand{\bibinfo}[2]{#2}
\providecommand{\BIBentrySTDinterwordspacing}{\spaceskip=0pt\relax}
\providecommand{\BIBentryALTinterwordstretchfactor}{4}
\providecommand{\BIBentryALTinterwordspacing}{\spaceskip=\fontdimen2\font plus
\BIBentryALTinterwordstretchfactor\fontdimen3\font minus
  \fontdimen4\font\relax}
\providecommand{\BIBforeignlanguage}[2]{{%
\expandafter\ifx\csname l@#1\endcsname\relax
\typeout{** WARNING: IEEEtran.bst: No hyphenation pattern has been}%
\typeout{** loaded for the language `#1'. Using the pattern for}%
\typeout{** the default language instead.}%
\else
\language=\csname l@#1\endcsname
\fi
#2}}
\providecommand{\BIBdecl}{\relax}
\BIBdecl

\bibitem{bahdanau2016end}
D.~Bahdanau, J.~Chorowski, D.~Serdyuk, P.~Brakel, and Y.~Bengio, ``End-to-end
  attention-based large vocabulary speech recognition,'' in \emph{2016 IEEE
  International Conference on Acoustics, Speech and Signal Processing
  (ICASSP)}.\hskip 1em plus 0.5em minus 0.4em\relax IEEE, 2016, pp. 4945--4949.

\bibitem{karita2019comparative}
S.~Karita, N.~Chen, T.~Hayashi, T.~Hori, H.~Inaguma, Z.~Jiang, M.~Someki,
  N.~E.~Y. Soplin, R.~Yamamoto, X.~Wang \emph{et~al.}, ``A comparative study on
  transformer vs rnn in speech applications,'' in \emph{2019 IEEE Automatic
  Speech Recognition and Understanding Workshop (ASRU)}.\hskip 1em plus 0.5em
  minus 0.4em\relax IEEE, 2019, pp. 449--456.

\bibitem{baevski2020wav2vec}
A.~Baevski, Y.~Zhou, A.~Mohamed, and M.~Auli, ``wav2vec 2.0: A framework for
  self-supervised learning of speech representations,'' \emph{Advances in
  Neural Information Processing Systems}, vol.~33, 2020.

\bibitem{cho2012segmentation}
E.~Cho, J.~Niehues, and A.~Waibel, ``Segmentation and punctuation prediction in
  speech language translation using a monolingual translation system,'' in
  \emph{International Workshop on Spoken Language Translation (IWSLT)}, 2012.

\bibitem{peitz2011modeling}
S.~Peitz, M.~Freitag, A.~Mauser, and H.~Ney, ``Modeling punctuation prediction
  as machine translation,'' in \emph{International Workshop on Spoken Language
  Translation (IWSLT)}, 2011.

\bibitem{nguyen2020improving}
T.~B. Nguyen, Q.~M. Nguyen, T.~T.~H. Nguyen, Q.~T. Do, and C.~M. Luong,
  ``Improving vietnamese named entity recognition from speech using word
  capitalization and punctuation recovery models,'' in \emph{Proc. Interspeech
  2020}, 2020, pp. 4263--4267.

\bibitem{gravano2009restoring}
A.~Gravano, M.~Jansche, and M.~Bacchiani, ``Restoring punctuation and
  capitalization in transcribed speech,'' in \emph{2009 IEEE International
  Conference on Acoustics, Speech and Signal Processing}.\hskip 1em plus 0.5em
  minus 0.4em\relax IEEE, 2009, pp. 4741--4744.

\bibitem{akita2006sentence}
Y.~Akita, M.~Saikou, H.~Nanjo, and T.~Kawahara, ``Sentence boundary detection
  of spontaneous japanese using statistical language model and support vector
  machines,'' in \emph{Proc. Interspeech 2006}, 2006, p. 1370.

\bibitem{lu2010better}
W.~Lu and H.~T. Ng, ``Better punctuation prediction with dynamic conditional
  random fields,'' in \emph{Proceedings of the 2010 conference on empirical
  methods in natural language processing}, 2010, pp. 177--186.

\bibitem{akita2011automatic}
Y.~Akita and T.~Kawahara, ``Automatic comma insertion of lecture transcripts
  based on multiple annotations,'' in \emph{Proc. Interspeech 2011}, 2011, pp.
  2889--2892.

\bibitem{hochreiter1997long}
S.~Hochreiter and J.~Schmidhuber, ``Long short-term memory,'' \emph{Neural
  computation}, vol.~9, no.~8, pp. 1735--1780, 1997.

\bibitem{vaswani2017attention}
A.~Vaswani, N.~Shazeer, N.~Parmar, J.~Uszkoreit, L.~Jones, A.~N. Gomez,
  L.~Kaiser, and I.~Polosukhin, ``Attention is all you need,'' in
  \emph{Advances in Neural Information Processing Systems (NeurIPS)}, 2017, pp.
  5998--6008.

\bibitem{tilk2015lstm}
O.~Tilk and T.~Alum{\"a}e, ``Lstm for punctuation restoration in speech
  transcripts,'' in \emph{Sixteenth annual conference of the international
  speech communication association}, 2015.

\bibitem{yi2019self}
J.~Yi and J.~Tao, ``Self-attention based model for punctuation prediction using
  word and speech embeddings,'' in \emph{ICASSP 2019 IEEE International
  Conference on Acoustics, Speech and Signal Processing (ICASSP)}.\hskip 1em
  plus 0.5em minus 0.4em\relax IEEE, 2019, pp. 7270--7274.

\bibitem{devlin2019bert}
J.~Devlin, M.-W. Chang, K.~Lee, and K.~Toutanova, ``Bert: Pre-training of deep
  bidirectional transformers for language understanding,'' in \emph{Proceedings
  of the 2019 Conference of the North American Chapter of the Association for
  Computational Linguistics: Human Language Technologies, Volume 1 (Long and
  Short Papers)}, 2019, pp. 4171--4186.

\bibitem{makhija2019transfer}
K.~Makhija, T.-N. Ho, and E.-S. Chng, ``Transfer learning for punctuation
  prediction,'' in \emph{2019 Asia-Pacific Signal and Information Processing
  Association Annual Summit and Conference (APSIPA ASC)}.\hskip 1em plus 0.5em
  minus 0.4em\relax IEEE, 2019, pp. 268--273.

\bibitem{yi2020adversarial}
J.~Yi, J.~Tao, Y.~Bai, Z.~Tian, and C.~Fan, ``Adversarial transfer learning for
  punctuation restoration,'' \emph{arXiv preprint arXiv:2004.00248}, 2020.

\bibitem{che2016punctuation}
X.~Che, C.~Wang, H.~Yang, and C.~Meinel, ``Punctuation prediction for
  unsegmented transcript based on word vector,'' in \emph{Proceedings of the
  Tenth International Conference on Language Resources and Evaluation
  (LREC'16)}, 2016, pp. 654--658.

\bibitem{liu2019roberta}
Y.~Liu, M.~Ott, N.~Goyal, J.~Du, M.~Joshi, D.~Chen, O.~Levy, M.~Lewis,
  L.~Zettlemoyer, and V.~Stoyanov, ``Roberta: A robustly optimized bert
  pretraining approach,'' \emph{arXiv preprint arXiv:1907.11692}, 2019.

\bibitem{clark2019electra}
K.~Clark, M.-T. Luong, Q.~V. Le, and C.~D. Manning, ``Electra: Pre-training
  text encoders as discriminators rather than generators,'' in
  \emph{International Conference on Learning Representations}, 2019.

\bibitem{courtland2020efficient}
M.~Courtland, A.~Faulkner, and G.~McElvain, ``Efficient automatic punctuation
  restoration using bidirectional transformers with robust inference,'' in
  \emph{Proceedings of the 17th International Conference on Spoken Language
  Translation}, 2020, pp. 272--279.

\bibitem{chen21d_interspeech}
Q.~Chen, W.~Wang, M.~Chen, and Q.~Zhang, ``{Discriminative Self-Training for
  Punctuation Prediction},'' in \emph{Proc. Interspeech 2021}, 2021, pp.
  771--775.

\bibitem{he2019streaming}
Y.~He, T.~N. Sainath, R.~Prabhavalkar, I.~McGraw, R.~Alvarez, D.~Zhao,
  D.~Rybach, A.~Kannan, Y.~Wu, R.~Pang \emph{et~al.}, ``Streaming end-to-end
  speech recognition for mobile devices,'' in \emph{ICASSP 2019-2019 IEEE
  International Conference on Acoustics, Speech and Signal Processing
  (ICASSP)}.\hskip 1em plus 0.5em minus 0.4em\relax IEEE, 2019, pp. 6381--6385.

\bibitem{graves2006connectionist}
A.~Graves, S.~Fern{\'a}ndez, F.~Gomez, and J.~Schmidhuber, ``Connectionist
  temporal classification: labelling unsegmented sequence data with recurrent
  neural networks,'' in \emph{Proceedings of International Conference on
  Machine Learning (ICML)}.\hskip 1em plus 0.5em minus 0.4em\relax PMLR, 2006,
  pp. 369--376.

\bibitem{christensen2001punctuation}
H.~Christensen, Y.~Gotoh, and S.~Renals, ``Punctuation annotation using
  statistical prosody models,'' in \emph{ISCA Tutorial and Research Workshop
  (ITRW) on Prosody in Speech Recognition and Understanding}, 2001.

\bibitem{sunkara2020multimodal}
M.~Sunkara, S.~Ronanki, D.~Bekal, S.~Bodapati, and K.~Kirchhoff, ``Multimodal
  semi-supervised learning framework for punctuation prediction in
  conversational speech,'' in \emph{Proc. Interspeech 2020}, 2020, pp.
  4911--4915.

\bibitem{fernandez2007sequence}
S.~Fern{\'a}ndez, A.~Graves, and J.~Schmidhuber, ``Sequence labelling in
  structured domains with hierarchical recurrent neural networks,'' in
  \emph{Proceedings of the 20th International Joint Conference on Artificial
  Intelligence, IJCAI 2007}, 2007.

\bibitem{sanabria2018hierarchical}
R.~Sanabria and F.~Metze, ``Hierarchical multitask learning with ctc,'' in
  \emph{2018 IEEE Spoken Language Technology Workshop (SLT)}, 2018, pp.
  485--490.

\bibitem{lee2021intermediate}
J.~Lee and S.~Watanabe, ``Intermediate loss regularization for {CTC-based}
  speech recognition,'' in \emph{2021 IEEE International Conference on
  Acoustics, Speech and Signal Processing (ICASSP)}.\hskip 1em plus 0.5em minus
  0.4em\relax IEEE, 2021.

\bibitem{Caseiro2020}
\BIBentryALTinterwordspacing
D.~Caseiro, P.~Rondon, Q.-N.~L. The, and P.~Aleksic, ``{Mixed Case Contextual
  ASR Using Capitalization Masks},'' in \emph{Proc. Interspeech 2020}, 2020,
  pp. 686--690. [Online]. Available:
  \url{http://dx.doi.org/10.21437/Interspeech.2020-2367}
\BIBentrySTDinterwordspacing

\bibitem{mimura2021endtoend}
M.~Mimura, S.~Sakai, and T.~Kawahara, ``An end-to-end model from speech to
  clean transcript for parliamentary meetings,'' in \emph{APSIPA}, 2021.

\bibitem{di2019must}
M.~A. Di~Gangi, R.~Cattoni, L.~Bentivogli, M.~Negri, and M.~Turchi, ``Must-c: a
  multilingual speech translation corpus,'' in \emph{2019 Conference of the
  North American Chapter of the Association for Computational Linguistics:
  Human Language Technologies}.\hskip 1em plus 0.5em minus 0.4em\relax
  Association for Computational Linguistics, 2019, pp. 2012--2017.

\bibitem{park2019specaugment}
D.~S. Park, W.~Chan, Y.~Zhang, C.-C. Chiu, B.~Zoph, E.~D. Cubuk, and Q.~V. Le,
  ``Specaugment: A simple data augmentation method for automatic speech
  recognition,'' \emph{Proc. Interspeech 2019}, pp. 2613--2617, 2019.

\bibitem{kudo2018sentencepiece}
T.~Kudo and J.~Richardson, ``Sentencepiece: A simple and language independent
  subword tokenizer and detokenizer for neural text processing,'' in
  \emph{Proceedings of the 2018 Conference on Empirical Methods in Natural
  Language Processing (EMNLP)}, 2018, pp. 66--71.

\bibitem{wolf-etal-2020-transformers}
\BIBentryALTinterwordspacing
T.~Wolf, L.~Debut, V.~Sanh, J.~Chaumond, C.~Delangue, A.~Moi, P.~Cistac,
  T.~Rault, R.~Louf, M.~Funtowicz, J.~Davison, S.~Shleifer, P.~von Platen,
  C.~Ma, Y.~Jernite, J.~Plu, C.~Xu, T.~Le~Scao, S.~Gugger, M.~Drame, Q.~Lhoest,
  and A.~Rush, ``Transformers: State-of-the-art natural language processing,''
  in \emph{Proceedings of the 2020 Conference on Empirical Methods in Natural
  Language Processing: System Demonstrations}.\hskip 1em plus 0.5em minus
  0.4em\relax Online: Association for Computational Linguistics, Oct. 2020, pp.
  38--45. [Online]. Available:
  \url{https://aclanthology.org/2020.emnlp-demos.6}
\BIBentrySTDinterwordspacing

\bibitem{kingma2015adam}
D.~P. Kingma and J.~Ba, ``Adam: A method for stochastic optimization,'' in
  \emph{3rd International Conference on Learning Representations, ICLR 2015,
  San Diego, CA, USA, May 7-9, 2015, Conference Track Proceedings}, 2015.

\end{thebibliography}

\end{document}